\documentclass[iop]{emulateapj}

\usepackage{amssymb}
\usepackage{natbib}
\usepackage{graphicx}

\def\beqn{\vspace{2mm} \begin{eqnarray}}
\def\eeqn{\vspace{2mm} \end{eqnarray}}
\newcounter{parentequation}

\def\kms{{\rm \thinspace km \thinspace s}^{-1}}

\def\Msun{\hbox{$\rm\thinspace M_{\odot}$}}
\def\pc{{\rm\thinspace pc}}     
       
\def\yr{{\rm\thinspace yr}}
\def\kyr{{\rm\thinspace kyr}}
\def\Myr{{\rm\thinspace Myr}}
\def\Gyr{{\rm\thinspace Gyr}}
\def\K{{\rm\thinspace K}}
\def\sdens{\thinspace\mathrm{cm}^{-2}}
\def\ndens{\thinspace\mathrm{cm}^{-3}}
\def\dens{\thinspace\mathrm{g}\thinspace\mathrm{cm}^{-3}} 
\def\AU{\thinspace\mathrm{AU}}
\def\Alrad{${\rm^{26}Al}$}
\def\Al{${\rm^{27}Al}$}
\def\Mg{${\rm^{26}Mg}$}
\def\Alratio{${\rm^{26}Al/^{27}Al}$}

\shorttitle{Supernova Triggered Formation and Enrichment of Our Solar System}
\shortauthors{M. Gritschneder, D.N.C. Lin, S.D. Murray,  Q.-Z. Yin, M.-N. Gong}

\begin{document}


\title{The Supernova Triggered Formation and Enrichment of Our Solar System}
\author{M. Gritschneder$^{1}$,D.N.C. Lin$^{1,2}$, S.D. Murray$^{3}$,
  Q.-Z. Yin$^{4}$, M.-N. Gong$^{5}$}
\affil{$^1$ Kavli Institute for Astronomy and Astrophysics, Peking University,
Yi He Yuan Lu 5, Hai Dian, 100871 Beijing, China \texttt{gritschneder@pku.edu.cn}}
\affil{$^{2}$Astronomy and Astrophysics Department, University of California, Santa Cruz,
  CA 95064, USA}
\affil{$^{3}$Lawrence Livermore National Laboratory, University of
  California, Livermore, CA 94550, USA}
\affil{$^{4}$ Department of Geology, University of California, Davis, CA
95616, USA}
\affil{$^{5}$Department of Physics, Tsinghua University, Hai Dian, 100084 Beijing, China}



\begin{abstract}
We investigate the enrichment of the pre-solar cloud core with short
lived radionuclides (SLRs), especially \Alrad.
The homogeneity and the surprisingly small spread in the ratio \Alratio\,
observed in the overwhelming majority of
calcium-aluminium-rich
inclusions (CAIs) in a vast variety of primitive chondritic meteorites
places strong constraints on the formation of the the
solar system. Freshly synthesized
radioactive \Alrad\,has to be included and well mixed within
$20\kyr$.
After discussing various scenarios including X-winds, AGB stars and
Wolf-Rayet stars,  we come to the conclusion that triggering the
collapse of a cold cloud core by a nearby supernova is the most promising scenario. 
We then narrow down the vast parameter space by considering the
pre-explosion survivability of such a clump as well as the
cross-section necessary for sufficient enrichment. 
We employ numerical simulations to address the mixing of
the radioactively enriched SN gas with the pre-existing gas and the
forced collapse within $20\kyr$.  We show that a cold clump of $10\Msun$ at a
distance of $5\pc$ can be sufficiently enriched in \Alrad\,and triggered into
collapse fast enough - within $18\kyr$ after encountering the supernova
shock - for a range of different metallicities and progenitor masses, even
if the enriched material is assumed to be distributed homogeneously in
the entire supernova bubble.
In summary, we envision an environment for the birth place of the
Solar System $4.567\Gyr$ ago
similar to the situation of the pillars in M16 nowadays, where molecular cloud
cores adjacent to an HII region will be hit by a supernova explosion in
the future. We show that the triggered collapse and formation of the Solar
System as well as the required enrichment with radioactive \Alrad\,
are possible in this scenario.
\end{abstract}




\maketitle

\section{Introduction}
The time-scale for the formation events of our Solar System can be
derived from the decay products of radioactive elements found in
meteorites. Short lived radionuclides (SLRs)
such as \Alrad\,, ${\rm^{41}Ca}$, ${\rm^{53}Mn}$ and ${\rm^{60}Fe}$
can be employed as high-precision and high-resolution chronometers due
to their short half-lives. These SLRs are found in a wide variety of
Solar System materials, including calcium-aluminium-rich
inclusions (CAIs) in primitive chondrites
\citep[e.g.][]{Lee:1976fk,Amelin:2002lr}.
The decay of \Alrad\, into \Mg\ in particular, with a
half-life of $\tau\approx0.7\Myr$, provides by far the highest
resolution measurements that mark the initial formation of the
proto-planetary disk \citep{Jacobsen:2008uq}. As \Alrad\,decays, the
ratio of  \Alratio\,at the time of condensation can be directly derived from the
present day ${\rm^{26}Mg}/{\rm^{24}Mg}$ and
${\rm^{27}Al}/{\rm^{24}Mg}$ measurements.
Furthermore, the spread of \Alratio\, among different CAIs gives the
precise time-span 
in which these CAIs formed.
While there is considerable spread among the CAIs towards lower ratios
of \Alratio\, due to remelting and thermal metamorphism, the upper
value, the so called 'canonical ratio', \citep{MacPherson:1995lr} is
now well established \citep[e.g.][]{Jacobsen:2008uq,Villeneuve:2009fj}.

The general picture we adopt here is that a certain amount of \Alrad\,
is injected in the nascent solar nebula and then gets incorporated
into the earliest formed CAIs as soon as the temperature drops below
the condensation temperature of CAI minerals. These CAIs are most
frequently found in CV-chondrites\footnote{CV-chondrites
  are a class of CAI-rich carbonaceous chondrites named after the
  Vigarano meteorite \citep{Dauphas:2011fk}}. Lower ratios of
\Alratio\, can then be explained by subsequent episodes of CAI
remelting or thermal disturbance, thereby explaining the heterogeneity below the canonical
value. 
Therefore, the CAIs found in chondrites represent the first known solid
objects that crystalized within our Solar System and can be used as an
anchor point to determine the formation time-scale of our
Solar System. Various
measurements of different CAIs by several research groups have not
only confirmed the canonical ratio of 
$(5.23\pm0.13)\times10^{-5}$, but also established a very small
spread (for a recent review see \citealt{Villeneuve:2009fj}). This
spread corresponds to an age range of less than $\simeq20\kyr$
\citep[e.g.][]{Jacobsen:2008uq}. 

In addition, the Mg-isotope composition appears to be fairly uniform
among bulk chondrites as well as Mars, Moon and Earth. This is a
strong indication that initially Al- and Mg-isotopes are distributed
quite homogeneously ($\pm 10\%$) in the proto-planetary disk
\citep{Thrane:2006fk,Villeneuve:2009fj,Boss:2011lr}. The extremely small time-span
together with the highly homogeneous mixing of isotopes poses a severe
challenge for theoretical models on the formation of our Solar System

Various theoretical scenarios for the formation of the Solar System
have been discussed. Shortly after the
discovery of SLRs, it was proposed that 
they were injected by a nearby massive star. This can happen either
via a supernova explosion \citep{Cameron:1977yq} or by the strong
winds of a Wolf-Rayet star \citep{Arnould:1997vn}. Another possibility
would be the in situ enrichment inside the disk by the non-thermal
activity of the Sun \citep{Shu:2001fk}. 
\citet{Wasserburg:1995fj} have proposed an asymptotic giant branch
(AGB) star as an possible source of SLRs in the Solar System 
(see also \citealt{Trigo-Rodriguez:2009kx}).

Up to now, none of these models have been able to explain the
extremely small
range of CAI condensation ages (the spread of $\approx 20\kyr$). In this 
paper, we discuss several possible solutions to this problem. 
First, we show that most of these scenarios are not
compatible with an almost coeval formation of CAIs. We then
go on to perform numerical simulations on the promising scenario of 
a type IIa supernova as the trigger of our Solar System.

\section{Physical Processes and Feasibility of Solutions}
\label{phys_proc}
In this section we discuss the physical processes in
different scenarios and their viability. For a detailed discussion of
these scenarios, including an assessment of their likelihood see
the recent review by \citet{Adams:2010lr}.

As mentioned before, the canonical value of \Alratio\, is established
to be $(5.23\pm0.13)\times10^{-5}$.
Assuming an total mass for \Al\, of
${M_{\rm^{27}Al}}=7.1\times10^{-5}\Msun$ \citep[e.g.][]{Lodders:2003fk} 
we can convert the observed ratio into a total mass of
${M_{\rm^{26}Al}^{\rm canonical}}\approx3.71\times10^{-9}\Msun$ of
radioactive 
\Alrad, 
which has to be present in the early Solar System.

\subsection{In situ Enrichment}
One solution to the incorporation of \Alrad\, into CAIs was
proposed by \citet{Shu:2001fk}. Young stars can undergo the so-called
X-wind phase. In this phase the combination of magnetic fields,
inflowing material and stellar outflows can form a wind
between the proto-star and the proto-planetary disk
\citep{Shu:1997qy}. 

SRLs are produced in this scenario by solar energetic particle
irradiation of dust and gas near the proto-Sun \citep{Shu:2001fk} and
incorporated into CAIs. This mechanism can explain the ratio of
${\rm^{10}Be/^{9}Be}=8.8\pm0.6\times10^{-4}$ inferred from
measurements of ${\rm^{10}B}$, the decay product of ${\rm^{10}Be}$
\citep{McKeegan:2000lr}. Another way to explain this ratio was
proposed by \citet{Desch:2004fk}. They calculate the contribution of
${\rm^{10}Be}$ trapped by galactic cosmic rays to the collapsing
molecular cloud core and conclude that a large amount, if not
all, ${\rm^{10}Be}$ can be attributed to cosmic rays.

For the purpose of this study it is sufficient to note that this
mechanism can not explain the observed amount of \Alrad\, due to the
following reasons. First of all, \Alrad\,and ${\rm^{10}Be}$ do not
correlate well in CAIs \citep[e.g.][]{Marhas:2002qy}. In addition, SLRs produced by this
mechanism are intrinsically expected to be highly
heterogeneous. Variations in their relative abundances would reflect
the local energetic particle environment and 
their episodic nature of production. This is in contradiction to the
observed homogeneity ($\pm10\%$) of \Alrad.
Furthermore, the amount of \Alrad\, which can be produced by this
mechanism is at least an order of magnitude too low to explain the
canonical value \citep{Duprat:2007fj,Duprat:2008uq}.

\subsection{External Enrichment by AGB stars}
Another source of \Alrad\, are AGB stars. Every star of
$\sim0.8-8\Msun$ \citep {Herwig:2005lr} undergoes this phase when
helium fusion in the core is complete and helium shell burning
begins. In this phase, the star loses most of its envelope in stellar
winds, which could in term enrich the proto-solar
cloud\citep[e.g.][]{Wasserburg:1995fj}. This stage is at the end of a 
stellar life. However, AGB stars are not present in star forming regions and
hence an encounter is very unlikely \citep{Kastner:1994lr}. 
In addition, the total amount of \Alrad\, requires a massive 
AGB-star and the short enrichment-timescale requires a brief AGB-phase.

\subsection{External Enrichment by Wolf-Rayet stars}
\label{WRstars}
An interesting possibility was proposed by \cite{Tatischeff:2010uq}
who followed up on an earlier idea by \cite{Arnould:1997vn}. Here, a run-away
Wolf-Rayet star that gets ejected from its parental cluster of massive
stars sweeps up a smaller shell while traveling into the ISM. This
shell gets enriched by material from the strong stellar winds of the
Wolf-Rayet star. As soon as this star explodes in a supernova it
sweeps up an even denser shell which expands rapidly. After
$\simeq10\kyr$ the shell starts to cool and a cold core can start to
form due to dynamical and gravitational instabilities which
subsequently undergoes collapse. During this collapse phase the the
surface density of the core increases with time
\citep{Whitworth:1994kx,Heitsch:2008fj}. Since this theory assumes
that the \Alratio\,ratio increases as the surface density of the clump
increases, the core becomes enriched sufficiently to explain the observed values in CAIs after
$10^5-10^6\yr$. However, there is no physical reason why this ratio should
increase as the core density rises. The initial ratio within the shell
should be conserved and only modified by radioactive decay
(the decay is taken into account by \citealt{Tatischeff:2010uq}).
 
A similar scenario was proposed by \cite{Gaidos:2009fk}. Here, an
entire molecular cloud ($M>10^5\Msun$) is enriched in \Alrad\,by a
nearby star forming region. After $\sim4-5\Myr$ this cloud begins
to form stars, including the Sun. During this stage, the Solar
System is enriched with heavier SLRs. This scenario can explain the
different abundances of SLRs (see \S \ref{discussion}). However, it
requires a precise chain of events. In addition, this theory assumes
continuous enrichment which is at odds with the small age spread of
CAIs (see below). 

\subsection{Constraints on continuous enrichment}
\label{contin_enrich}
All three scenarios described above are based on the assumption of a 
continuous enrichment.  In that case, the characteristic time scale
for the enrichment with radioactive isotopes is much longer than the
brief duration ($<20\kyr$)  of the CAI formation episode which is
inferred from the meteoritic data.

In external enrichment models, sufficient radioactive isotopes can only
be intercepted by a tenuous and extended solar-system progenitor cloud 
prior to its collapse.  But the free fall time scale for such a cloud is generally
$\sim 100\kyr$ or longer.  If the central star and disk accrete material from 
this parent envelope on such a time scale, dust and grains would be processed
with different rates at different locations and in different stages.  During this
entire phase, the assembly and formation of the CAIs would be possible, 
leading to inhomogeneities and an age-spread of $\sim 100\kyr$.

This is inconsistent with the observationally inferred 
homogeneity of Mg- and Al-isotopes during the formation epoch of CAIs. 
Here, we make a clear distinction and differentiate the initial
formation epoch of CAIs from later epochs when some CAIs subsequently experienced a
complex thermal history within the solar nebula and metamorphism on
their host meteorite parent asteroids extending over $0.1-3\Myr$ 
\citep[e.g.][]{MacPherson:1995lr,Hsu:2000fk,Young:2005kx,MacPherson:2007uq}.
One possible mechanism to homogenize the isotope's distribution and 
reset the clock for grain condensation and CAI formation is to 
heat the infalling gas above the condensation temperature for CAIs of
$\sim1500-1800$K \citep[e.g.][]{Grossman:2002yq,Lodders:2003fk}.  During
the infall, accretion 
and shocks can dissipate sufficient energy and heat up the disk to
this temperature close to the 
central star (within $\sim 0.1$ AU).  In order for the infalling gas
to arrive at
this location and form a compact disk, it must carry very little
angular momentum.
However, turbulent angular momentum transport, induced either by
gravitational or magneto-rotational instabilities, would lead to mass
diffusion.  When the disk
spreads beyond an AU, its mid-plane cools
\citep{Ruden:1986lr,Garaud:2007fk} well below the condensation
temperature of the CAIs. If the infalling gas
has a modest amount of angular momentum, it would form more extended
disks with shock temperatures below $10^3$K at a distance outside 1 AU
\citep{Walch:2009qy}.  In either limit, condensation of the elements and the
formation of CAI's would continue until infall is terminated.

A possible solution to reconcile with CAI's brief formation epoch is
to assume that all
infalling gas was heated above their condensation temperature of
$\sim1800$K and then
cooled within $20\kyr$.  Such a
scenario would require a very compact disk and therefore a violent,
external cause, i.e. a supernova, to 
trigger and drive the collapse of the entire progenitor cloud which
formed the Solar System. As shown in Appendix A, the
subsequent growth of CAIs would be possible on a short enough
time-scale in such a disk. 

\subsection{External Enrichment by a Type II Supernova}
\begin{figure}
  \centering{
   \plotone{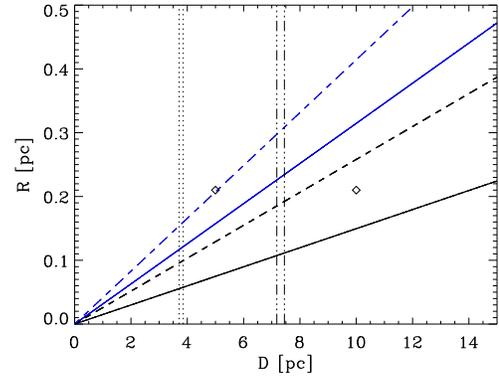}
  \caption{The parameter space. Plotted is the distance D from the
    exploding star versus the radius R of the pre-solar clump. The
    diagonal lines give the required ratio for the enrichment. Black
    is for a $40\Msun$ star, blue for a $20\Msun$ star, solid is solar
    metallicity with Z=0.02, dashed is Z=0.004. The vertical lines are
    the radius of the HII-region for an O5V star after 3Myr and a B0V
    star after 10Myr, respectively - dotted is for a cold medium of
    $5\times10^{4}\ndens$, dash-dotted for $5\times10^{3}\ndens$. The
    diamonds represent the parameters chosen for the simulations in \S
    \ref{simulations}.
\label{param_space}}}
\end{figure}
A frequently discussed possibility for SLRs is the external enrichment by
a type IIa supernova (SN IIa). \citet{Huss:2009lr} estimate that a supernova
with a progenitor of $20-60\Msun$ can produce the required
SLRs. From the simple assumption that
the cross-section of the enriched region has to be high enough to
intercept enough \Alrad\, we can derive a size-distance relation
from geometrical considerations:
\begin{equation}
\label{eq_crosssection}
R=\sqrt{\frac{4\cdot M_{\rm Al^{26}_{\odot}}}{M_{\rm Al^{26}_{SN}}}} D
\end{equation}
Where $R$ is the radius of the intercepting cloud or disk and $D$ is
the distance from the supernova. $M_{\rm Al^{26}_{\odot}}$ and
  $M_{\rm Al^{26}_{SN}}$ are the amount of \Alrad\, required in the
  Solar System and produced in the supernova, respectively. This
  equation is of course simplified, e.g. neglecting radioactive decay
  as the time-scales are very short, as well as assuming a perfect
  mixing efficiency, i.e. all \Alrad\, intercepted is assumed to be
  mixed into the Solar System. 
\begin{table}
\begin{center}
\begin{tabular}{lcc}
&${M_{\rm Al^{26}_{SN}}}[\Msun]$& \\
\hline
Mass $[\Msun]$ & Z=0.02 & Z=0.004 \\
\hline
20 & $1.5\times 10^{-5}$ & $8.65\times 10^{-6}$ \\ 
40 & $6.64\times 10^{-5}$ & $2.23\times 10^{-5}$ \\
\end{tabular}
\caption{Typical \Alrad\, yields in $\Msun$ for different supernova progenitor masses and
  metallicities taken from \cite{Nomoto:2006lr}.
\label{yields_tab}}
\end{center}
\end{table}

\subsubsection{Enrichment in the Disk Stage}
From the values in Table \ref{yields_tab} one can calculate that at
solar metallicity a $40\Msun$ progenitor supernova could 
deposit enough \Alrad\, in a $1000\AU$ face-on disk at $D=0.3\pc$ or a
$20\Msun$ progenitor enough \Alrad\, at 
$D=0.15\pc$. That is very close to the hot, ionizing star.
Although a proto-stellar disk could survive in this violent environment
\citep{Adams:2010lr}, there are additional timing
issues. For example \citet{Gounelle:2008fk} estimate the probability that the
proto-solar system is young enough ($<1\Myr$), able to survive
photo-evaporation and close enough to get sufficiently enriched
($D<0.3\pc$) to be less than $10^{-3}$ in the most favorable
case. Based on the required young age of the system,
\citet{Williams:2007yq} argue for a very massive supernova progenitor,
which in turn leads to a short life-time, i.e. an early supernova. As
very massive stars are only found in massive clusters, they conclude
that the likelihood is $<1\%$.

\cite{Ouellette:2005kx,Ouellette:2009fk,Ouellette:2010qy} suggested an
intriguing possibility where the  
disk is further away from the supernova ($\sim 2\pc$) and the
enrichment is achieved via the injection of already formed grains into
the pre-solar disk ('aerogel' model). To achieve the high observed
yield, the supernova gas has to be highly clumpy and the
clumps have to be highly enriched in SLRs - to this possibility the authors estimate a
probability of $\sim10^{-3}-10^{-2}$. There are, however,
  additional uncertainties. Recent models \citep[e.g.][]{Ida:2004lr} suggest that the
minimum mass solar nebula might be more massive than
their assumed $0.01\Msun$, which would require an even more efficient
enrichment. Future investigation on the clumpiness of SN-shock fronts
are desirable (see \S \ref{discussion}).
A remaining issue in this picture is the question why this scenario
does not result in multiple enrichments of the early Solar
System. After the first enrichment by a supernova, which leads to the
canonical value of \Alratio, it is likely that several more lighter and
therefore more long-lived stars in the cluster undergo subsequent
supernova. This would result in several enrichment episodes during the
entire lifetime of the disk ($>10\Myr$).

\subsubsection{Enrichment in the Cloud Core Stage}
The enrichment of the pre-solar cloud core by a supernova has been
  suggested for a long time \citep{Cameron:1977yq,Cameron:1995qy}. Recently, 
\citet{Boss:2008uq,Boss:2010qy} and \citet{Boss:2010lr} investigated the implosion
and enrichment of a $1\Msun$ cloud core in detail. They assume the
cloud core to be more than $10\pc$ away from the
progenitor. Therefore, the supernova front has slowed down
considerably by the time it reached the pre-solar cloud. In addition,
it has a finite thickness. In their models, the shock speed ranges
from $v_{\rm{S}}=1-100\kms$ \citep{Boss:2010qy}, the shock thickness
from $\delta_{\rm{S}}=3\times10^{-4}-3\times10^{-2}\pc$ and the shock
density from $\rho_{\rm{S}}=3.6\times10^{-21}-2.4\times10^{-17}\dens$ \citep{Boss:2010lr}. They find 
that with certain initial conditions the enrichment is possible in
their model. For example a thin ($\delta_{\rm{S}}=3\times10^{-4}$),
dense( $\rho_{\rm{S}}=6\times10^{-18}\dens$) shock at a rather low 
velocity ($v_{\rm{S}}=20\kms$) and therefore far away from the source can enrich the
clump sufficiently. This can be easily
understood as they assume that the enriched supernova material is
completely within the shock. Therefore, as the geometrical dilution
rises, the shock has to be thinner to contain a sufficient amount of
supernova material. In addition, the time-scale of the collapse is
longer ($t>40\kyr$) than the value inferred from CAIs in this
scenario. Again, this can be attributed to the slower supernova
shock further away from the progenitor. Nevertheless, this is an
encouraging development and we think this scenario is the most
promising.  

Bearing the geometrical dilution in mind (cf Fig. \ref{param_space})
we revisit this scenario but place the pre-solar clump closer to the
supernova. Therefore, the shock velocities are much higher (see \S
\ref{IC}). We also take the more conservative assumption that
the material is spread out in the hot phase or at least in a ring at the border
within the supernova explosion bubble (see \S \ref{sim5pc} and
\ref{sim10pc}). Furthermore, we do not assume the shock already formed
a density enhancement. Instead, we setup the 
supernova blast wave in the velocities under the assumption of a thin shock.
As the evolution is expected to be much more violent, we investigate
the collapse of a heavier ($10\Msun$) cloud core (see \S\ref{IC}). To
assess the feasibility of this initial condition, we address the aspect
of survival of the pre-solar cloud (\S \ref{survival}).

A slightly different approach would be the enrichment in the cloud
stage without triggering the collapse. Then, the cold cloud would
collapse in an 'isolated' manner at a later stage. 
However, there are clear short-comings of this scenario. To allow for
the tight spread of only $\simeq20\kyr$, the collapse would have to
lead to a stage, where the entire disk is heated above
$\sim1800\K$ and then homogeneously cools to lower temperatures
within $20\kyr$ (see also \S \ref{contin_enrich}).
To achieve the
required heating from the gravitational energy the accretion disk
would have to be $<0.5\AU$. At this scale, the disk could then cool
coherently, the CAI could condense and later migrate outwards to
explain the homogeneous \Alratio\, in CAIs in different meteorites. 

However, isolated collapse leads to larger, cooler disks. Thus, a
triggering event will be required to force the collapse to a small
scale. It is straightforward to assume that this triggering event
was the same event as the enrichment, i.e. the supernova shock. Thus,
we investigate the triggered case in more detail.

\section{Model and Simulations}
\label{simulations}
In this section, we address the parameter space by accessing the
enrichment, survivability and existence of the
pre-solar cloud core 
in or adjacent to the HII region of the massive star. 
We investigate the mixing and the collapse
probabilities by employing numerical simulations. 

In Fig. \ref{param_space} we plot the distance of the clump to the
progenitor star versus the size of the clump. From the geometrical enrichment
cross-section discussed before, we can derive a first
constraint. Taking the yields in Table \ref{yields_tab} into account
we plot Eq. \ref{eq_crosssection}. These diagonal lines 
denote the maximum enrichment
efficiency. Even if all the material intersected by the initial cloud
ends up in the Solar System ($100\%$ efficiency), the initial condition has to be
above these lines to explain the ratio inferred from CAIs.

\subsection{Survival}
\label{survival}
The most obvious proof for the existence of cold cloud cores in the
proximity of massive star comes from observations. In the pillars
in M16 \citep[e.g.][]{Hester:1996fk} \cite{McCaughrean:2002uq} find
several cold cores in almost every region of the pillars. These
pillars are about $2\pc$ away from the ionizing sources, the age of the
region is estimated to be $1-2\Myr$ \citep{Hester:1996fk}. This agrees well with our
assumption of structures being present at $5\pc$ after $3\Myr$ (see below). 

From a theoretical point of view, recent simulations on the formation
and evolution of HII-regions show the ubiquity of these structures
\citep[e.g.][]{Mellema:2006yq,Krumholz:2007kx,Gritschneder:2009uq,Arthur:2011qy}.
These simulations focus mostly on the ionization of turbulent molecular
clouds. As shown in \cite{Gritschneder:2010fj}, the average time
evolution of the front position is still well approximated by an analytic
solution as given in \cite{Shu:1991vn}:
\begin{equation}
\label{x_front}
R(t)=R_\mathrm{s}\left(1+\frac{7}{4}\frac{c_\mathrm{s,hot}}{R_\mathrm{s}}(t-t_0)\right)^\frac{4}{7},
\end{equation}
where $R_\mathrm{s}$ is the initial Stroemgren radius\footnote{The
  Stroemgren radius is the radius which can be immediately ionized by
  an ionizing source, without taking any hydrodynamic evolution into
  account \citep{Stromgren:1939lr}.} and $c_\mathrm{s,hot}$ is the
sound speed of the hot, ionized gas. Using this equation, we can
estimate the size of the HII-region. To investigate the most extreme
cases, we assume a short-lived $40\Msun$-star with a very short life-time
of $3\Myr$ as well as a $20\Msun$-star with a life-time
of $10\Myr$ \citep[e.g.][]{Hurley:2000qy} in different density
regions. The resulting radii are depicted by the vertical lines in
Fig. \ref{param_space}. Surviving clumps should be to the right of
theses lines. Note that we assume a relatively high density in the
neutral medium before the ignition of the star in order to
reflect the birth environment of massive stars. \citet{Myers:2009uq}
suggest that the birth of stellar groups is
associated with hubs of column densities greater than $10^{22}\sdens$, for
example in Taurus \citep{Goldsmith:2008qy}, which corresponds to the
assumed densities here. In addition, the high densities have to be
present only in the direct surrounding of the massive stars, as the
further evolution given in Eq. \ref{x_front} only depends on the initial
Stroemgren radius (and the sound speed in the hot gas). Altogether,
these values are chosen to reflect our current understanding of
massive star formation. In addition, Fig. 1 gives the averaged front
position, the pillars in the observations as well as the simulations
can be a few pc further inside, pertruding into the ionized bubble. 

\subsection{Collapse Timescale and Mixing in the Gas Phase}
\begin{figure*}
  \centering{
\includegraphics[height=9.8cm]{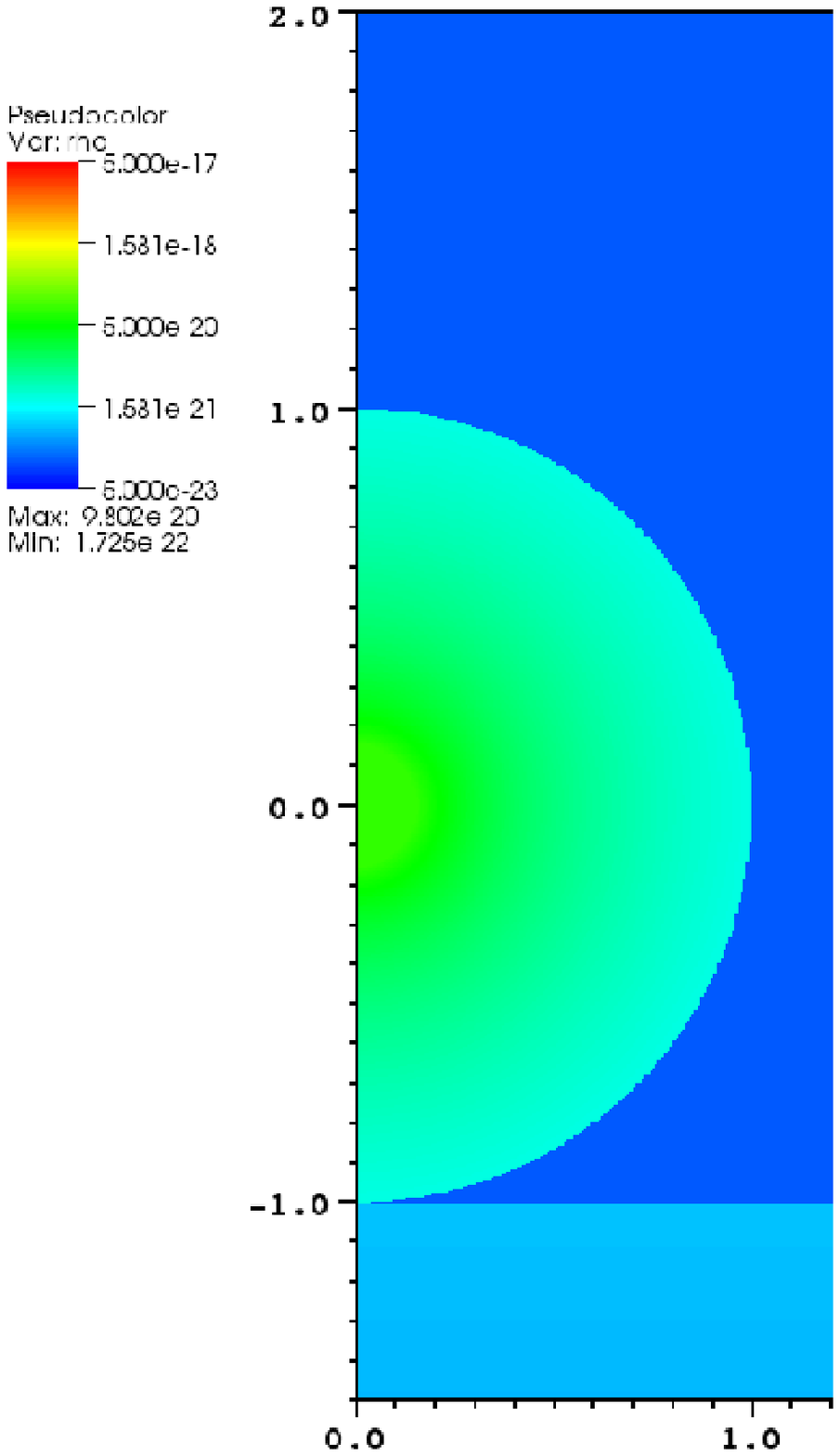}
\includegraphics[height=9.8cm]{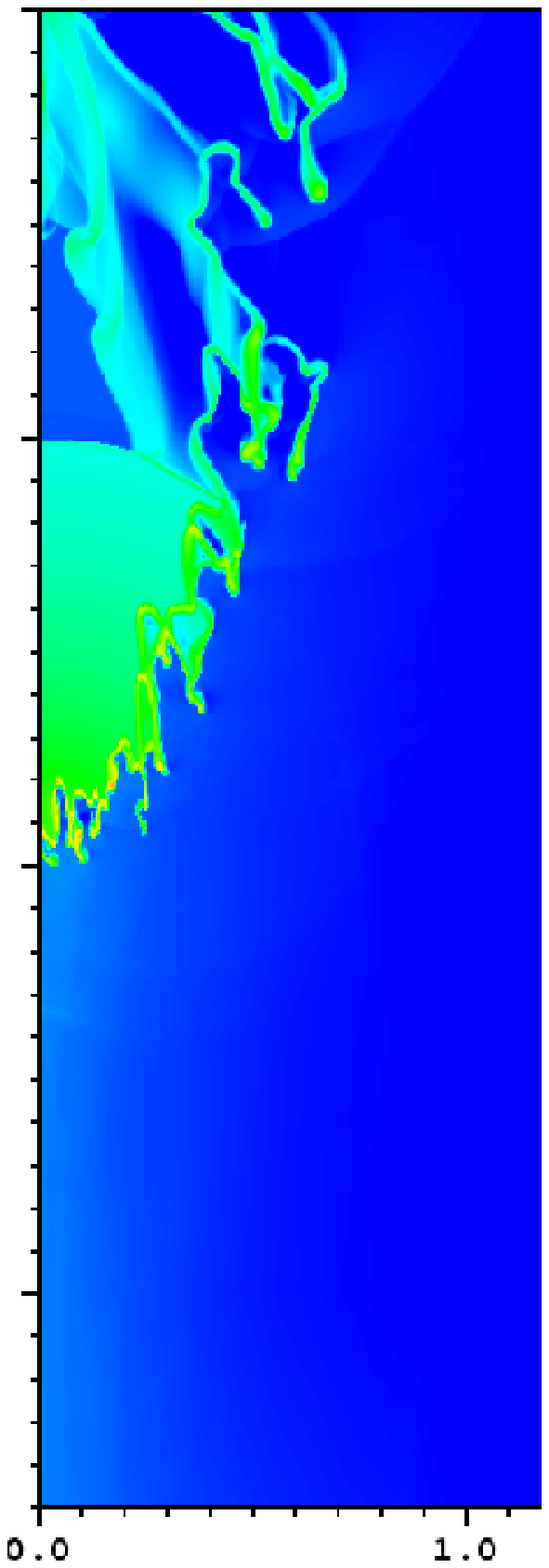}
\includegraphics[height=9.8cm]{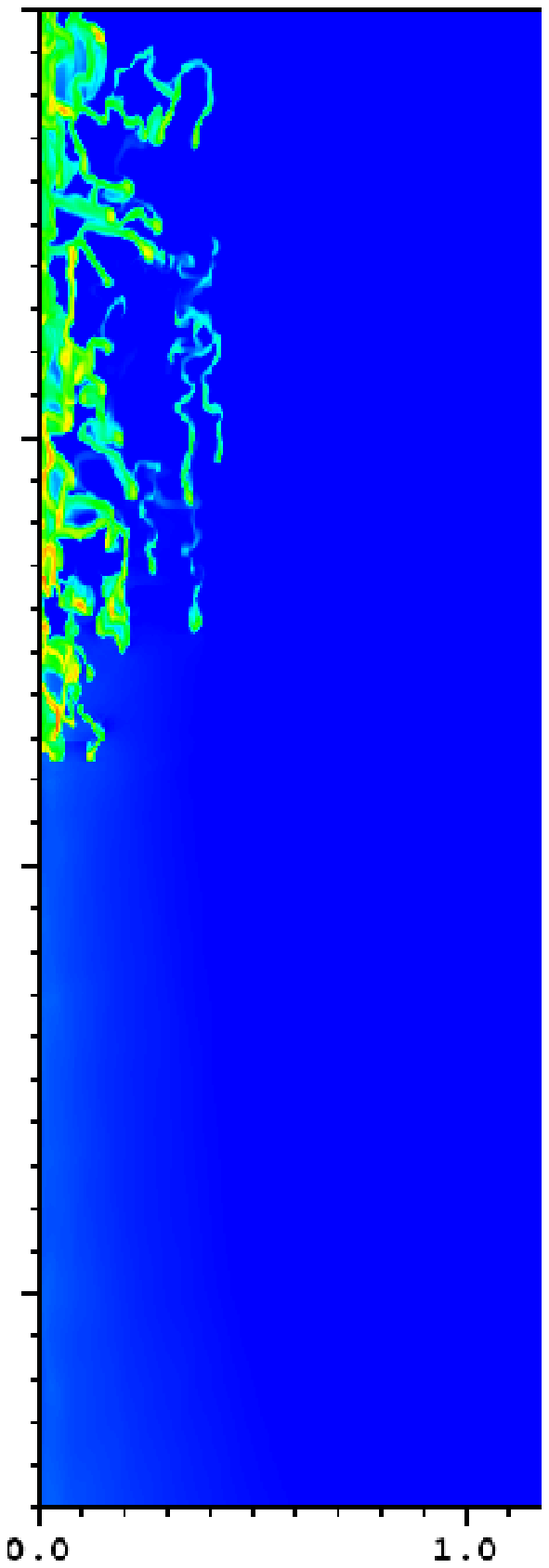}
 \caption{The time evolution of case I. Color coded is the density at
   $t=0\kyr$, $t=4.16\kyr$ and $t=8.33\kyr$. The length scale is
   given in units of the radius of the initial cold core ($R_0=0.21\pc$).\label{time_evol}}}
\end{figure*}
The remaining issues are whether the gas from the supernova can be
efficiently mixed with the gas from the pre-existing core, whether
collapse occurs and if so, whether it is fast enough. To test these
questions, we employ numerical simulations.

\subsubsection{Initial Conditions}
\label{IC}
Because the main focus lies on mixing as well as cooling, we employ the
numerical code COSMOS.  The non-relativistic hydrodynamics of COSMOS are
discussed in \citep{Anninos:2003qy}. For the models described below, the
code is run with the default hydrodynamics solver, which is a total
variation diminishing (TVD) approximate Riemann solver and a third-order
Runge-Kutta time-marching scheme \citep{Shu:1988ys}.  Because of the
importance of cooling, the code is run in internal energy mode.  Shocks are
handled with a zone-centered scalar artificial viscosity.  Optically thin
heating and cooling are included, using equilibrium cooling curves
\citep{Dalgarno:1972rt,Boehringer:1989vn}.  Cooling by metals
assumes solar metallicity.  Mixing is followed by the use of a passive
tracer, whose value is initialized to be proportional to the gas density
behind the blast-wave.  The problems are run in two dimensions.  Cylindrical
symmetry is imposed by taking advantage of the relativistic capability of
COSMOS, and running the problems with a cylindrical metric.

The problem region runs from -5 to 10 cloud radii in the direction of the
shock, and from 0 to 4 cloud radii in the transverse direction.  The large
radial size is adopted so as to prevent any boundary affects from altering
the cloud evolution.  Due to the motion of the cloud after shock passage, we
use a uniform grid spacing in the direction of the incoming shock, with 4096
zones along the axis, or 274 zones across the cloud radius.  In order to
enhance resolution of the collapsed cloud, geometrically increasing zone
spacing is used in the radial direction.  A total of 512 zones are included
radially, with a ratio of 1.005 in their spacing, giving a minimum zone size
of 0.0017 times the cloud radius.

We set up an molecular cloud core in isolation, which is going to be
hit by the supernova shock-wave. We approximate the core by an
marginally stable Bonnor-Ebert-Sphere \citep{Bonnor:1956lr}, the
solution of the Lane-Emden equation for hydrostatic equilibrium
\begin{equation}
  \frac{1}{r^2}\frac{d}{dr}(\frac{r^2}{\rho}\frac{dp}{dr})=-4\pi G\rho,
\end{equation}
where gravity and hydrodynamical pressure balance each other. 
We assume an initial temperature of $T_\mathrm{cold}=20\K$ and a
central number density of
$n_\mathrm{c}=5.7\times 10^4\ndens$, which gives a cloud core of mass
$M=10\Msun$ with a radius $R_0=0.21\pc$. The sphere is embedded
in a warm gas in pressure equilibrium with the sphere at a temperature
$T_\mathrm{warm}=792\K$. This is realistic, as the core is embedded in
the interstellar medium close to an HII-region. 
As this is an idealized case and the main focus of this study is the
enrichment with SRLs, we
don't take any rotational or turbulent motion
\citep[e.g.][]{Walch:2010fk} inside the cloud core into account.
While this initial rotation might be important in the later phases
  of the collapse - especially in determining the final disk size -
  its influence will be very small in the initial phase here, as 
  the rotational velocities are orders of magnitude smaller than the
  shock speeds. In addition, the total angular momentum carried by a
  minimum mass nebula ($0.02\Msun$) out to $30\AU$ corresponds to that of
  $1\Msun$ of material at $0.6\AU$.  Unless there is a substantial angular
  momentum loss since the initial collapse to the setup of a minimum
  mass nebula, the initial cloud must carry very little angular
  momentum. In such a low-angular momentum cloud, rotation is unlikely
  to inhibit the collapse until the size of the cloud is well below the
resolution of our simulations and the dynamical time scale is much
shorter than the $20\kyr$ of interest.

The supernova shock-wave at the distances considered here ($5-10\pc$)
has not yet reached the snow-plow phase. So the shock is still in the
energy dominated phase and can be approximated by a Sedov-Taylor blast
wave \citep{Taylor:1950qy,Sedov:1959uq}. Following
\citet{Woosley:1995fj} we assume an energy release of
$E=10^{51}$erg for core-collapse supernovae with progenitor
stars in the range of $20-40\Msun$. Considering the parameter space
(\S \ref{survival}), we perform two simulations, placing the cloud
core at a distance $D=5\pc$ (case I) and $D=10\pc$ (case II), respectively,
from the supernova. We evaluate the Sedov solution for the given
distances. In case I, the results are that the blast wave reaches the core after
$\simeq8.5\kyr$ with a post-shock speed of $v_\mathrm{s}=276\kms$. The
sound-speed in the post-shock gas is $c_\mathrm{s}=129\kms$. In case
II the cold cloud core is reached after $\simeq48.2\kyr$ with
$v_\mathrm{s}=97.5\kms$ and $c_\mathrm{s}=45.4\kms$.

\subsubsection{Case I}
\label{sim5pc}
In the first case, we place the cold pre-solar core at a distance
$D=5\pc$ from the supernova. Fig. \ref{time_evol} shows the time
evolution of the density in this simulation. The shock wave is
propagating from the bottom to the
top. As it can be clearly seen, the shock wave encompasses the cold
core rapidly. Various hydrodynamical instabilities occur at the
interface. After $t=4.15\kyr$ the front has already passed the
core. The part of the front which hit the clump continues to interact
with the pre-existing core, leading to two fronts wrapping around
it. As these fronts collide, the material gets mixed with the
pre-existing cold gas and is driven into collapse. After
$t=8.33\kyr$ the central region is already at a very high density
($\rho_\mathrm{max}=3\times10^{-15}\dens$), the computation gets
expensive and we terminate it.

\begin{figure}
  \centering{
\plotone{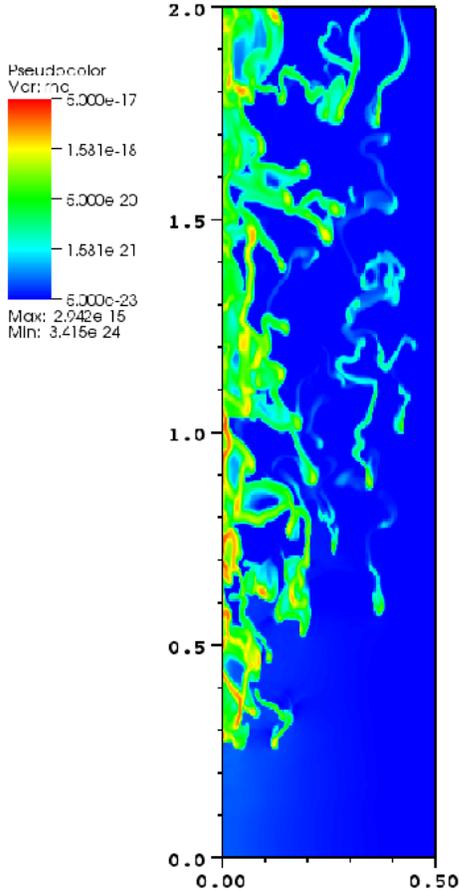}
  \caption{Zoom in into the final stage of Case I at
    $t=8.33\kyr$. Color coded is the density, the contours denote the
    gravitational potential.  The length scale is
   given in units of the radius of the initial cold core
   ($R_0=0.21\pc$). As can be clearly seen the densest 
    regions will undergo gravitational collapse.\label{zoom5pc}}}
\end{figure}

\begin{figure}
  \centering{
\plotone{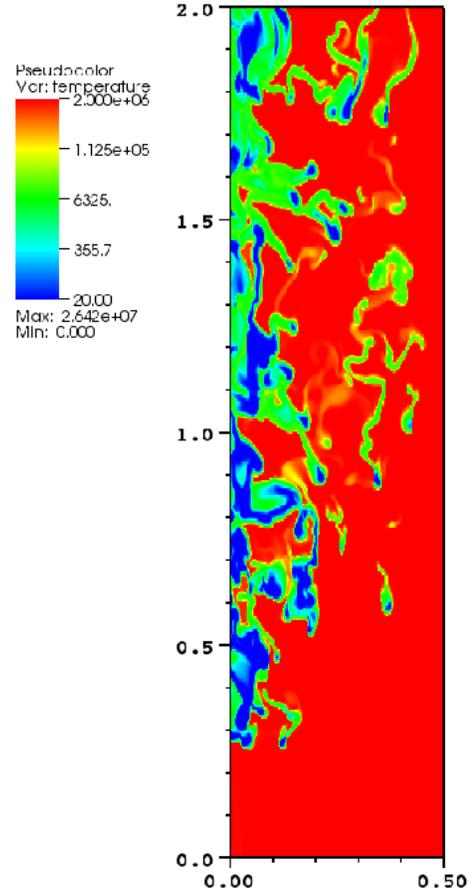}
  \caption{The temperature distribution at the final stage at
    $t=8.33\kyr$. The blue are cold regions, red are hot regions. The length scale is
   given in units of the radius of the initial cold core ($R_0=0.21\pc$). The dense regions
    are very cold, indicating gravitational collapse.\label{zoom5pcT}}}
\end{figure}

Fig. \ref{zoom5pc} shows a zoom-in into the density at the final
snapshot. In Fig. \ref{zoom5pcT} we show the temperature
distribution in this region.  As one can already see, the central
region is cold. We define the central region as the cold, dense
  material between $0.25$ and $0.75$ on the y-axis of
  Fig. \ref{zoom5pcT}, corresponding to an elongated filament of
  $\approx0.1\pc$ diameter. The mass in this region below a temperature of
  $20\K$ is ${M_{\rm core}}\simeq0.13\Msun$, 
the average density is
$\bar{\rho}=1\times10^{-17}\dens$. This gives a
Jeans-mass\footnote{The Jeans-mass is calculated with a mean
  molecular weight of $\mu=2.5$, i.e. assuming that molecular hydrogen
and helium are the dominant species.} of $\simeq0.08\Msun$. The core
is twice as heavy and therefore will definitely undergo collapse. 
This collapse will proceed at least with the free-fall
time-scale, which is $t_\mathrm{ff}\approx 20\kyr$ at this
average density. More likely, it will proceed faster as the central
density is two orders of magnitude higher and since the material is
already moving inwards. In summary, it is probably safe to say that the
further collapse to a stage below $1800\K$, where the CAIs form, takes
less than $10\kyr$.

The mass is still lower than the solar value, but given the fact that
cooling and accretion will still happen this value can
increase. Although, given from the mass available in the surrounding we do not
estimate it to reach more than $0.2\Msun$ in total. In
addition, very strong inflows or colliding remnants of the shock are
still present, flowing towards the centre with velocities of
$\simeq10\kms$. These will lead to an enhanced, fast collapse likely
force the material in a small central region, which favors the simultaneous
crystallization of CAIs.

The other main focus here is the mixing of the enriched hot supernova
gas with the cold pre-existing clump. To follow this evolution, we
include a tracer in the post-shock gas inside the shock-front. The
ratio of \Alrad\,to the total gas mass within the supernova bubble depends on several
factors. First of all, different progenitor masses give different
yields. Secondly, the exact distribution of the enriched gas
within the supernova bubble remains an open question. 

We investigate
two cases here. The worst case, where the supernova enrichment is
fully mixed with the entire material in the hot bubble, i.e. the
$20\Msun$ of total supernova yield are diluted with $\sim1000\Msun$
within the bubble\footnote{To estimate the dilution we assume an average density
  of $n=100\ndens$ within the supernova bubble from here on. This is a
high value, but still realistic (for a discussion see \S \ref{discussion}).}. The
second case is more optimistic, assuming that the enriched material
stays close to the blast-wave. We assume it to be within the outer $10\%$ of
the radius of the bubble, leading to a dilution by only
$\sim350\Msun$ of pre-supernova material.

\begin{figure}
  \centering{
 \plotone{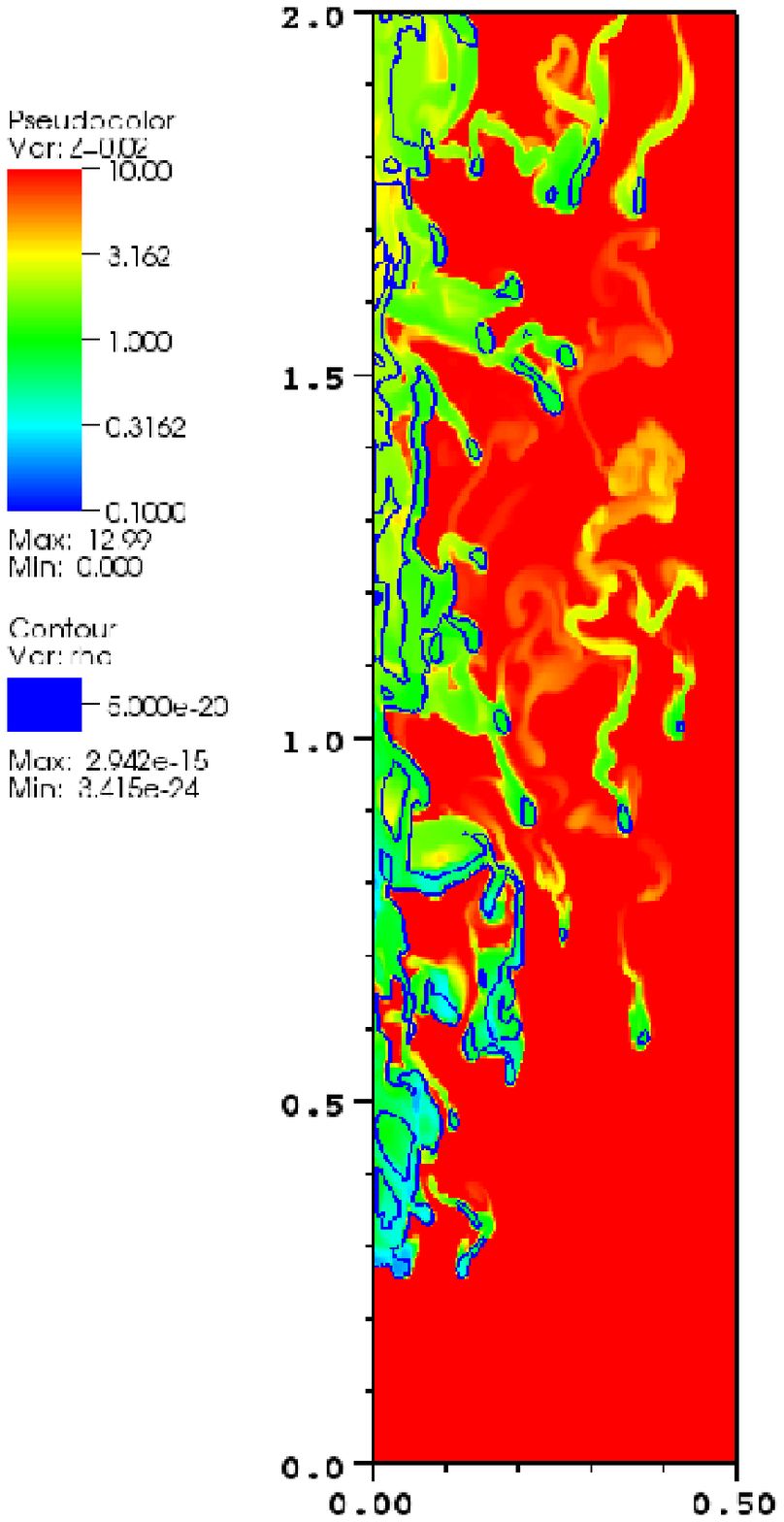}
 \caption{The tracer field, which follows the supernova gas at
   $t=8.33\kyr$. Color coded is the ratio \Alratio, the required value
   as inferred from CAI measurements is unity (green).  The length scale is
   given in units of the radius of the initial cold core
   ($R_0=0.21\pc$). The color
   coding here is for a $40\Msun$ progenitor at $5\pc$ distance and
   the supernova gas getting fully mixed within the entire supernova
   bubble. The contours show the density distribution. As it can be
   seen the collapsing pre-solar gas gets sufficiently enriched in
   this case.\label{mixing}}} 
\end{figure}

However, even in the worst case of complete mixing, the enrichment is
sufficient for a $40\Msun$ progenitor star at solar metallicity. This is shown in
Fig. \ref{mixing}. Color coded is the ratio of \Alrad\,to total mass,
normalized to the required value of $\simeq3.71\times10^{-9}$. Since
in the collapsing regions the ratio is of order unity or higher, we
conclude that the gas phase is sufficiently enriched to allow for the
formation of the observed CAIs later on. 

In Table \ref{results_tab} we list the outcome of different
  scenarios. We give ${M_{\rm^{26}Al}^{\rm sim}}$, the total mass of
\Alrad\, inside the final collapsing core, as well as
  the dilution $f_{0}$ (i.e. the fraction of supernova material
  injected into the final core) and the injection efficiency  $f_{\rm i}$
  as defined in Eq. 
  \ref{eq_enrich}. Finally, we list the resulting enrichment for a
  collapsing core scaled to the solar ratio
\begin{equation}
f_{\odot} = \frac{{M_{\rm^{26}Al}^{\rm sim}}}{M_{\rm core}}\frac
{M_{\odot}}{M_{\rm^{26}Al}^{\rm canonical}}.
\end{equation}
$f_{\odot}$ can be interpreted as the success of the model, i.e. value
of 1.0 corresponds to the observed \Alratio\, in the Solar System.
 
Our results differ depending on the assumed distribution of enriched
material inside the supernova bubble and on the progenitor mass. As can be seen,
in several cases the enrichment is sufficient, if the cloud core is
placed at $D=5\pc$ from 
the supernova initially. Note that the dilution $f_0$ is
systematically lower than the values of $f_0\approx (0.5-6)\times
10^{-4}$ as given by \citet{Takigawa:2008lr}. At the same time
the mass in the cold core is only $0.13\Msun$ in our case, explaining
that difference of one order in magnitude. In addition, we assume a
high number density inside the supernova bubble ($n=100\ndens$),
which leads to a very strong dilution. Thus, these estimates are
very conservative and can be viewed as a worst case scenario
enrichment.

\begin{figure}
  \centering{
 \plotone{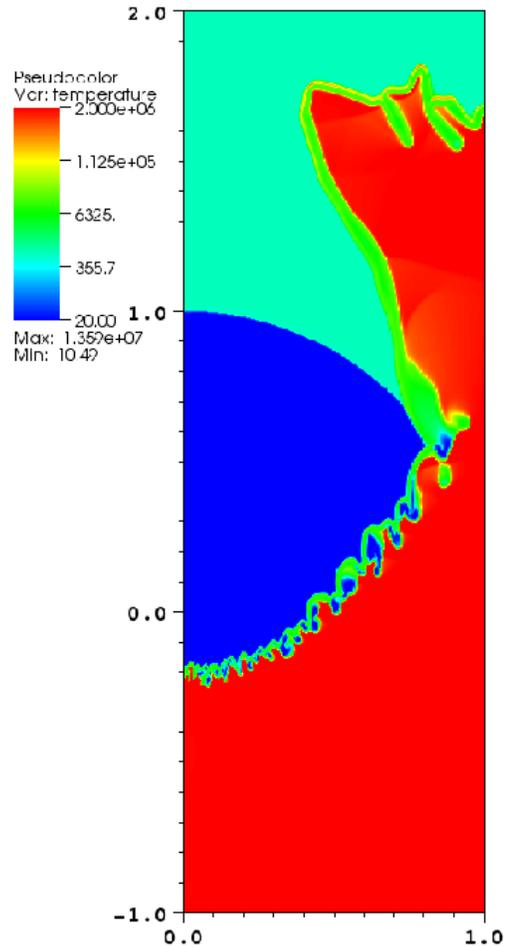}
 \caption{ The temperature distribution at a much earlier stage
   ($t=1.25\kyr$). Blue is cold, red is hot.  The length scale is
   given in units of the radius of the initial cold core
   ($R_0=0.21\pc$). As can be seen, the
   enriched gas cools behind/within the shock-front and only later
   gets mixed with the pre-existing cold gas. The mixing happens
   between the two cold phases, rendering it more efficient. \label{mixingT}}}
\end{figure}

To further understand the reason for this very efficient mixing, it is
useful to take a closer look at the earlier stage of the
simulation. In Fig. \ref{mixingT} 
we show the temperature distribution at the intermediate stage
($t=1.25\kyr$). As can be clearly seen the material is already cooling
behind the shock front. This cold, enriched clumps are mixing very
efficiently with the pre-existing core. Therefore, a precise
prescription of cooling processes is vital to investigate enrichment scenarios. 

\begin{table*}
\begin{center}
\begin{tabular}{lccccccc}
\hline
Distance & Mass & Metallicity & SNe mixing & ${M_{\rm^{26}Al}^{\rm sim}}[\Msun]$ & $f_{0}$ & $f_{\rm i}$ &$f_{\odot}$ \\
\hline
$5\pc$ & $40\Msun$ & $Z=0.02$ & shell & $1.6\times10^{-9}$ &
$2.3\times10^{-5}$ & $5.3\times10^{-2}$ & 3.34 \\
& & & complete & $4.4\times10^{-10}$ &  $6.6\times10^{-6}$ & $1.5\times10^{-2}$ & 0.91 \\ 
& & $Z=0.004$ & shell & $5.4\times10^{-10}$ & $2.3\times10^{-5}$ & $5.3\times10^{-2}$ &  1.1\\
 & &  & complete & $1.6\times10^{-10}$ & $7.0\times10^{-6}$ & $1.6\times10^{-2}$ &  0.32\\
& $20\Msun$ & $Z=0.02$ & shell & $3.6\times10^{-10}$ & $2.5\times10^{-5}$ & $5.6\times10^{-2}$ & 0.75\\
& & & complete & $1.0\times10^{-10}$ & $6.6\times10^{-6}$ & $1.5\times10^{-2}$ & 0.21\\
& & $Z=0.004$ & shell & $1.6\times10^{-10}$ & $2.5\times10^{-5}$& $5.6\times10^{-2}$ & 0.43 \\
 & &  & complete & $5.7\times10^{-11}$ & $6.5\times10^{-6}$ & $1.5\times10^{-2}$ & 0.11\\
\hline
$10\pc$ & $40\Msun$ & $Z=0.02$ & shell & $3.9\times10^{-12}$ & $5.8\times10^{-8}$& $5.2\times10^{-4}$ & 0.065  \\
& & & complete & $1.1\times10^{-12}$ & $1.6\times10^{-8}$ & $1.5\times10^{-4}$ & 0.017\\ 
& & $Z=0.004$ & shell & $1.3\times10^{-12}$ & $5.8\times10^{-8}$ & $5.2\times10^{-4}$ & 0.022\\
 & &  & complete & $3.8\times10^{-13}$ & $1.8\times10^{-8}$  & $1.6\times10^{-4}$ & 0.0064\\
& $20\Msun$ & $Z=0.02$ & shell & $8.7\times10^{-13}$ & $5.8\times10^{-8}$ & $5.2\times10^{-4}$ & 0.014\\
& & & complete & $2.4\times10^{-13}$ & $1.6\times10^{-8}$ & $1.5\times10^{-4}$ & 0.0041\\
& & $Z=0.004$ & shell & $5.1\times10^{-13}$ & $5.9\times10^{-8}$ & $5.4\times10^{-4}$ & 0.0085  \\
 & &  & complete & $1.4\times10^{-13}$ & $1.6\times10^{-8}$ & $1.4\times10^{-4}$ & 0.0023 \\
\end{tabular}
\caption{Outcome of different scenarios. We give the distance, mass,
  chosen metallicity and assumed SNe mixing. The corresponding yields
  are taken from \cite{Nomoto:2006lr}. We give ${M_{\rm^{26}Al}^{\rm sim}}$, the total mass of
  \Alrad\, inside the final core, as well as
  the dilution $f_{0}$ and the injection efficiency  $f_{\rm i}$. Finally we list the resulting enrichment for a
  collapsing core scaled up to the solar mass, i.e. a value of 1.0 or bigger
  stands for successful enrichment to the solar value.\label{results_tab}}
\end{center}
\end{table*}

\subsubsection{Case II}
\label{sim10pc}
In the second case, the cold cloud core is at a distance $D=10\pc$ and
is embedded in an ambient medium which is a factor of ten less dense
with a ten times higher temperature than case I. Therefore, the
ambient pressure is the same. 
Due to the longer distance the angular cross-section
is smaller, the enriched material is already more dispersed. In
addition, the shock has already lost more energy, i.e. the front speed
is slower. 

\begin{figure}
  \centering{
 \plotone{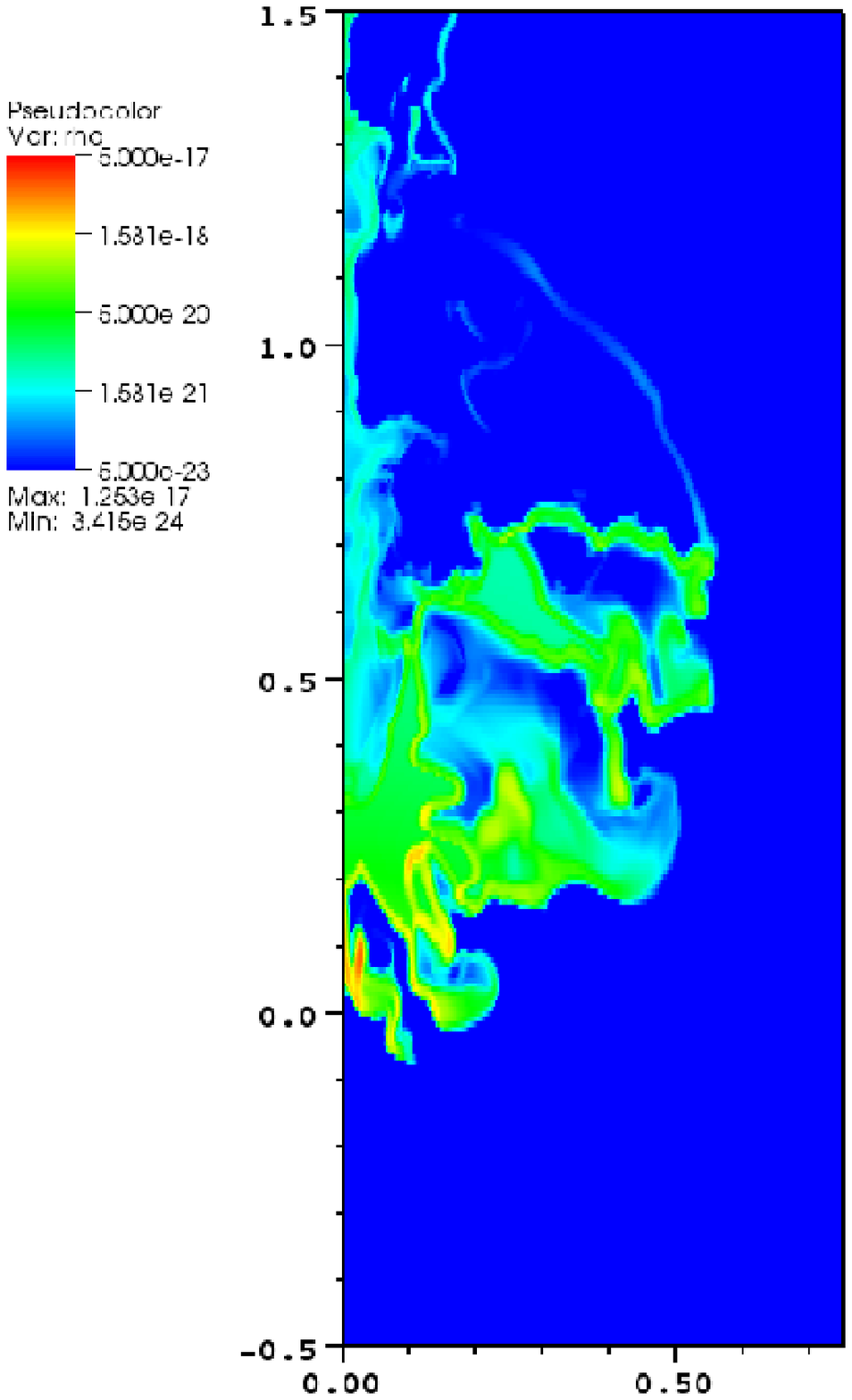}
  \caption{Zoom in into the final stage of Case II at
    $t=14.66\kyr$. Color coded is the density, the length scale is
    given in units of the radius of the initial cold core
    ($R_0=0.21\pc$). Not even the densest regions undergo 
    gravitational collapse.\label{zoom10pc}}}
\end{figure}

\begin{figure}
  \centering{
 \plotone{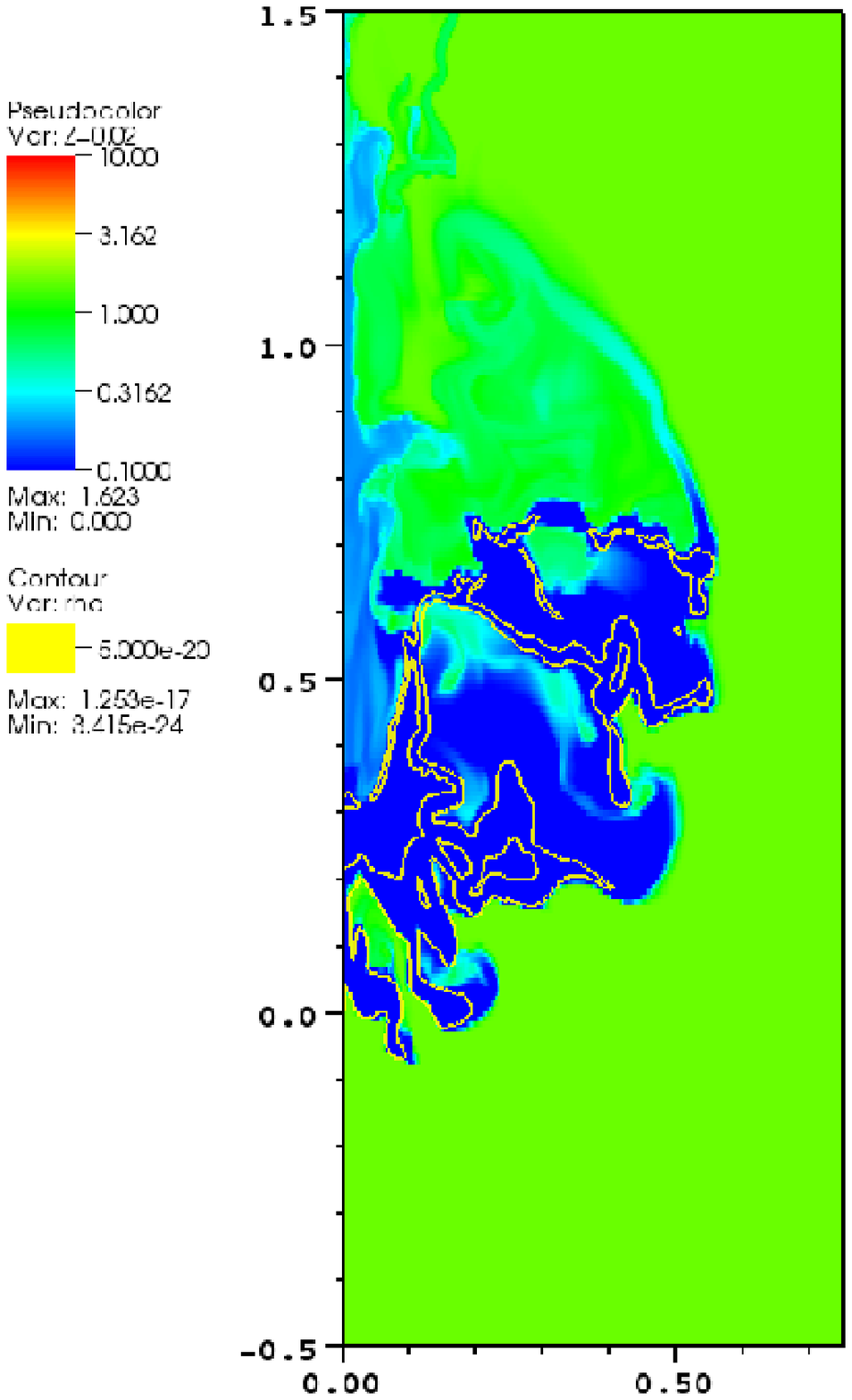}
  \caption{The tracer field, which follows the supernova gas for case II at
   $t=14.66\kyr$. Color coded is the ratio \Alratio, the required value
   as inferred from CAI measurements is unity (green). The length scale is
   given in units of the radius of the initial cold core ($R_0=0.21\pc$). The color
   coding here is for a $40\Msun$ progenitor at $10\pc$ distance and
   the supernova gas getting fully mixed within the entire supernova
   bubble. The contours show the density distribution. The only
   regions where the cold gas gets slightly enriched are the 
   colliding flanks in the back. Furthermore, no region undergoes
   gravitational collapse.\label{mix10pc}}}
\end{figure}

The effect of this can be seen in
Fig. \ref{zoom10pc}. After $t=14.6\kyr$ the cold core is still much less
deformed by the supernova and not driven into collapse.
Although the mixing seems to be very efficient, a closer look at
Fig. \ref{mix10pc} shows that the gravitational collapsing region,
a remnant of the original cloud core is not strongly enriched (the
blue region at $0.0-0.1$ cloud radii showing the highest density in
Fig \ref{zoom10pc}), whereas the enriched region
further away (the light blue region between $0.7-1.3$ cloud radii) shows
the effect of the two flanks of enriched supernova 
material colliding.

In addition, the mass in the cold region is much smaller, only
$0.01\Msun$, 
which is 30 times less than the Jeans mass at the average
density of $\bar{\rho}=7\times10^{-19}\dens$ in this
region. Therefore, the cloud will not collapse in this case. This is
due to supernova shock wave already being substantially weakened so
far away from the front. Of course it would be possible to start with a
more condensed core, but then the enrichment, which is already
difficult in this case (see below), becomes even more unlikely as the
radius of the core would be smaller (cf Eq. \ref{eq_crosssection}).

From Table \ref{results_tab} it can be seen that sufficient enrichment
is very unlikely in this scenario, especially since the enriched
material is now more diluted, as the volume of the supernova bubble
is bigger, i.e. the material produced by the supernova is now mixed
with $\sim10^4\Msun$ in the complete mixing case and with
$\sim2800\Msun$ in the shell case. 

Altogether the clump will neither collapse, nor be sufficiently
enriched in this case.

\subsection{Comparison of the enrichment in simulations with theoretical
  (geometrical) predictions}
\label{enrichment}
Here, we compare the predicted enrichment from the geometrical
cross-section (Eq. \ref{eq_crosssection} and Fig. \ref{param_space})
with the finally achieved enrichment.

From a geometrical point of view, the dilution $f_0$ is given as by the
radius of the initial sphere $R_0$ and the distance $D$ to the source as:
\begin{equation}
\label{eq_dilution}
f_{0\rm,geom}=\frac{R_0^2\pi}{4D^2\pi}.
\end{equation}
Thus,  $f_{0\rm,geom}$ is the amount of \Alrad\, intercepted geometrically by the
initial Bonnor Ebert sphere. $f_{0\rm,geom}=4.4\times 10^{-4}$ and
$f_{0\rm,geom}=1.1\times 10^{-4}$ for cases I and II, respectively.
The values are systematically higher than $f_0$ in Table
\ref{results_tab} as Eq. \ref{eq_dilution} assumes perfect mixing of the
supernova material with the background gas. However, it can be used to
determine an 
injection efficiency, which we can not define directly, as we do not
assume the material to be in the front initially.

Instead, we define the injection efficiency $f_{\rm i}$ as the ratio
of the finally incorporated mass to expected intercepted mass from
geometrical considerations
\begin{equation}
\label{eq_enrich}
f_{\rm i}=\frac{M_{\rm^{26}Al}^{\rm
    sim}}{f_{0\rm,geom}M_{\rm^{26}Al}^{\rm SNe}}=\frac{f_{0}}{f_{0\rm,geom}}.
\end{equation}
$f_{\rm i}$ would be unity if all intercepted material is included in the collapsing
core. We give the values in Table \ref{results_tab} as well. For case I
$f_{\rm i}\approx 0.01-0.05$, i.e. a few percent of the material 
crossing the cold sphere initially are incorporated in the collapsing
core finally.
In case II, the values are much lower,
$f_{\rm i}\approx0.0001-0.0006$. First, this is due to a larger amount of
un-enriched material in the bigger bubble. Second, the shock front is
much weaker in case II, and therefore the front does not mix as
efficient with the cloud core.

\section{Discussion}
\label{discussion}
The outcome of these simulations
  and the results given here are a first order proof of
  principle. From the current scenario it seems possible that every
  supernova with a progenitor of $20-40\Msun$ can produce a
  system with solar-like \Alrad\, abundances by triggering a nearby
  ($\approx5\pc$) cloud core
  into collapse. However, we only tested a small range of the
  parameter space. Future simulations are desirable to further probe
  the likelihood. A first issue would be that we assume the
  (idealized) analytical Sedov-Solution for the blast wave. A more
  realistic treatment for the entire evolution of the shock from the
  progenitor is required in the future. This is of course
  computationally very expensive. Nevertheless, it is important to
  test e.g. the fragmentation and cooling of the shock front during
  the $8\kyr$ before it hits the cloud core. Another area to improve would be
 to include more detailed effects. For example, the core could be initially
  rotating. We assume here that the high shock velocities dominate
  over the rotation, but this should be further investigated. Furthermore, we neglect magnetic
  fields. They are of course present and should be involved in various
  processes, e.g. the details of the shock front and especially later
  on in the disk formation and the removement of angular momentum. In
  the early stages, however, the kinetic energy of the shock plays the
  dominant role. Still, it is worthwhile to investigate this regime of the
  parameter space in future simulations.

A remaining issue is the production of heavy SLRs like ${\rm^{53}Mn}$
and  ${\rm^{60}Fe}$. \citet{Gounelle:2009uq} suggest that the observed
amount of ${\rm^{60}Fe}$ can be inherited from the ISM. Still, the
amount of galactic \Alrad\, is an order of magnitude lower than
in the Solar System \citep{Diehl:2006fj}. Thus, a different enrichment mechanism for
\Alrad\, is still needed. In addition, in the
supernova-enrichment scenario, ${\rm^{60}Fe}$ and especially
${\rm^{53}Mn}$ are over-produced by a factor of 10-100 relative to
their abundance in meteorites when compared to the lighter SLRs
${\rm^{41}Ca}$ and
\Alrad\,\citep[e.g.][]{Rauscher:2002lr,Gaidos:2009fk}. This is often
used as an argument against the supernova-enrichment
scenario. However, there are several possible 
solutions. First of all, the precise determination of abundances,
especially of radioactive isotopes in supernovae remains challenging. Second,
there are alternative models, e.g. fallback-mixing within
supernovae. Here, the innermost layers (commonly within the Si-burning
layer) fall back onto the star. Therefore, there are much less 
SLRs heavier than Si within the supernova shock \citep[e.g.][]{Takigawa:2008lr}. 
Third, as it is an overproduction,
not a lack of these SLRs, there could be different incorporation
efficiencies into CAIs and meteorites, which are not yet well
understood.

On a side note, the distribution of the supernova yield in the
  entire bubble versus the distribution of it in an outer rim gives
  different results (cf Table \ref{results_tab}). As the iron is
  produced in layers further inside it is
  valid to speculate that the different ratios might be due to an
  homogeneous distribution of ${\rm^{60}Fe}$ in the entire bubble,
  whereas \Alrad\, might be concentrated in an outer layer. This
  scenario is of course only valid if the initial distribution of
  elements in the supernova is conserved while the bubble expands
  to $5\pc$.

The different abundances of SLRs are often used to calibrate the delay
between the supernova and the formation of CAIs. Mostly, these calculations give
values of about a $\Myr$ (\citealt{Takigawa:2008lr}: $0.7\Myr$,
\citealt{Looney:2006lr}: $1.8\Myr$). This contradicts the fast formation time
required to produce the small spread observed in the CAIs.
In our opinion, the supernova yields of SLRs are at present not understood well
enough to derive meaningful delay time scales between
supernova and CAI formation. Especially, since
there might be different processes during the disk formation, which
could also explain the different abundances of SLRs. 

In our simulations we find that a distance $D\approx5\pc$ is optimal to keep
the balance of survivability for the pre-existing cold clump and the sufficient
enrichment. This is larger than previous estimates -
e.g. \cite{Looney:2006lr} derive a distance smaller than $4\pc$
from analytical estimates. The reason for this is the efficient mixing
as discussed in \S \ref{sim5pc} (see also Fig. \ref{mixingT}).

For the survivability of the pre-solar clump we assume a quite high
density and short lifetime for the HII-region. This leads to a small
size, i.e. a small distance for the pre-existing clump from the
O-star. The general picture would be the ionization front running into
the paternal filament, from which the O-star originally formed.
  On a related note, this leads to the assumed high density of
  the ambient surrounding of $n=100\ndens$ (according to the
  analytical evolution of an HII-region, see
  e.g. \citealt{Shu:1991vn}). This value is well within the observed
  range. \citet{Lefloch:2002lr} estimate the density in the
  Trifid Nebula to be $50\ndens$ whereas \citet{Rubin:2011lr} find values of
  $87-714\ndens$ and $74-1041\ndens$ for different lines in optical
  and infrared observations, respectively. In turn, the chosen value
  leads to a high density inside the supernova bubble, i.e. a very strong dilution of the
  enriched material. Thus, optimistic conditions for the survival
  lead to  very conservative or pessimistic assumptions for the
  dilution.
In addition, O/B-stars tend to form in bigger associations, therefore
there would be more ionizing flux than from a single
O-star. Furthermore, we neglect the 
effect of stellar winds. Altogether, the conditions for the survivability have
been chosen quite optimistically to have a cold core close enough to the O-star
to be sufficiently enriched ($D\approx5\pc$). Still, as the
observations around various HII-regions show, pillar like-structures
and therefore cold clumps  are observed frequently at these distances.

On a more technical side, these simulations are still only in 2D, so
some physical processes might not be well resolved. This might apply
to the gravitational collapse of two fronts colliding. However,
colliding fronts
of these speeds and densities have been shown to become
gravitationally unstable with full 3D simulations
\citep[e.g.][]{Gritschneder:2009lr}.
Another short-coming of 2D simulations might be the different
behavior of instabilities and mixing in 2D and 3D
\citep[e.g.][]{Stone:2007fk}. This should be less of an issue here, as
the mixing appears in the cold phase and is not solely due to 
hydrodynamical instabilities, which would be less well represented in
2D (see Fig. \ref{mixingT}). The mixing is mainly influenced by the
cooling, which can be well approximated in 2D.
Nevertheless, further simulations in full 3D would be desirable to address
this issues.

\section{Conclusions \& Outlook}
In this study, we focus on the hydrodynamical interaction of a
supernova with the pre-solar cloud core. A realistic supernova
shock wave is assumed to interact with the cloud. This leads to the
conclusion that it is possible to enrich a pre-existing clump
sufficiently with \Alrad\, to explain the observed ratio in the Solar
System. In order to survive the previous HII-region and still be
close enough to be sufficiently enriched and triggered into collapse,
it turns out that $D\approx5\pc$ is an ideal distance.

The previously stable core gets enriched and is triggered into
gravitational collapse within $t\approx8\kyr$ after the first
interaction with the supernova shock-wave. The further collapse
until the formation of the CAIs, which takes places at temperatures
below $1800\K$, can be estimated to be shorter than $\simeq10\kyr$.
Whether the collapse finally goes all the way down to the condensation
of the CAIs within $18\kyr$ or if there is an intermediate stage with
a hot disk where the pre-meteorite gas is well mixed remains to be
investigated. 
In either case, the CAIs will be able to grow fast enough to the
observed cm-sizes within the background density of a disk, as shown in
Appendix A.
The hot disk, however, can as well only be 
achieved by a very violent (i.e. supernova triggered) collapse. In addition, the
subsequent expansion and cooling of the disk will be rapid at these
high densities so it is preferable if the material is well mixed
before. 

On a side note, it is very interesting that the mixing occurs mainly
by material from the supernova shock, which already cooled. This
cooled portion interacts with the pre-existing cold gas in the
surrounding of the progenitor star, whereas
most of the supernova enriched material stays in the hot
component. Therefore, it might be possible to determine the efficiency
of enrichment by supernova feedback in the much larger context of
galaxies by investigating the cooling rates within supernova shocks. 

Altogether, the precise time and location of the grain formation
has to be determined by future investigations. In this work, we prove that it is possible to
  trigger a pre-existing stable core into collapse by a very nearby
  supernova ($D\approx5\pc$). In addition, we show for the first time
  that the collapsing core can be enriched sufficiently with SLRs to
  explain the observed abundances, even if the enriched material is
  not assumed to be in the shock initially, but instead distributed
  homogeneously within the entire supernova bubble.

\begin{acknowledgements}
We would like to thank the referees for valuable comments on the
manuscript. We thank Josh Wimpenny for carefully proof reading the
final manuscript.
M.G. acknowledges funding by the China National Postdoc Fund Grant No. 20100470108 and
the National Science Foundation of China Grant No. 11003001. 
D.N.C.L. acknowledges support by NASA grant NNX07AI88G, NNX08AL41G and
NNX08AM84G as well as the NSF grant AST-0908807.
Q.Z.Y. acknowledges NASA Cosmochemistry grant NNX08AG57G and Origins of Solar
Systems grant NNX09AC93G. The work by S.D.M. was performed under the
auspices of the Lawrence Livermore National Security, LLC under
contract No. DE-AC52-07NA27344. 
\end{acknowledgements}

\bibliographystyle{apj}
\bibliography{references}

\appendix

\section{Assembly and Growth of CAIs}
Assuming that Aluminum sticks perfectly to the grains, the growth rate $\dot{M}$  for a
particle of size $a$ at a certain radius within the disk with a
relative velocity $v_\mathrm{rel}$ is
\begin{equation}
\dot{M} = \pi a^2 \rho_m v_\mathrm{rel},
\end{equation}
where  $\rho_\mathrm{m}$ is the background density of the CAI-forming
material, e.g. \Al.
If we assume that, as the particle grows the density of the particle
$\rho_\mathrm{p}$ as well as $\rho_\mathrm{m}$
remain constant, we can write
\begin{equation}
4\pi a^2 \dot{a} \rho_\mathrm{p} = \pi a^2 \rho_\mathrm{m}
v_\mathrm{rel}
\end{equation}
and solving for $\dot{a}$
\begin{equation}
\dot{a} = \frac{\rho_\mathrm{m}}{4\rho_\mathrm{p}}v_\mathrm{rel}.
\end{equation}
If we ignore the migration of particle, that is, the particle stays in
a certain radius $r$ as it grows, the radius $a$ simply grows linearly
with time 
\begin{equation}
a = a_0 + \frac{\rho_\mathrm{m}}{4\rho_\mathrm{p}}v_\mathrm{rel}t =
a_0 + kt,
\end{equation}
where the growth rate $k = \frac{\rho_\mathrm{m}}{4\rho_\mathrm{p}}v_\mathrm{rel}$ is a function of radius.
Since we investigate a longer period of time ($t\sim20\kyr$), the initial particle size is
irrelevant, so we can assume $a_0 = 0$. Therefore,
\begin{equation}
\frac{\rho_\mathrm{m}}{4\rho_\mathrm{p}}v_\mathrm{rel}=\frac{a}{t}.
\end{equation}
To assemble cm-size objects within $20\kyr$, this gives:
\begin{equation}
\frac{\rho_\mathrm{m}}{4\rho_\mathrm{p}}v_\mathrm{rel}=1.6\times10^{-17}\kms.
\end{equation}
Assuming a density $\rho_\mathrm{p}=1\dens$ and a relative velocity of
$v_\mathrm{rel}=1\kms$, the resulting Aluminum background density has to be $\rho_\mathrm{m}=6.4\times10^{-17}\dens.$
With the total mass \Al\, of  ${M_{\rm^{27}Al}}=7.1\times10^{-5}\Msun$
\citep[e.g.][]{Lodders:2003fk} 
this leads to a required total background density of
\begin{equation}
\rho_\mathrm{disk}=9.0\times10^{-13}\dens,
\end{equation}
which is higher than the density in cloud cores, but can be easily
reached in proto-stellar and proto-planetary disks. 

On the one hand this is an optimistic scenario for the growth, since
we assume a perfect sticking efficiency. On the other hand, the 
injected supernova material most likely already contains $\sim0.1\mu$m-sized
grains \citep[e.g.][]{Nath:2008fr}. Furthermore, we only take \Al\,
into account here, whereas the CAIs are not formed by pure \Al.

\end{document}